\documentclass[twocolumn,showpacs,preprintnumbers,amsmath,amssymb]{revtex4}

\usepackage{graphicx}
\usepackage{dcolumn}
\usepackage{bm}

\begin{document}

\preprint{APS/123-QED}

\title{One-dimensional quantum random walks with two entangled coins}

\author{Chaobin Liu}
 \email{cliu@bowiestate.edu}
\author{Nelson Petulante}%
 \email{npetulante@bowiestate.edu}
\affiliation{%
Department of Mathematics, Bowie State University, Bowie, MD, 20715 USA\\
}%
\date{\today}

\begin{abstract}
We offer theoretical explanations for some recent observations in numerical simulations of quantum random walks (QRW). Specifically, in the case of a QRW on the line with one particle (walker) and two entangled coins, we explain the phenomenon, called ``localization", whereby the probability distribution of the walker's position is seen to exhibit a persistent major ``spike" (or ``peak") at the initial position and two other minor spikes which drift to infinity in either direction. Another interesting finding in connection with QRW's of this sort pertains to the limiting behavior of the position probability distribution. It is seen that the probability of finding the walker at any given location becomes eventually stationary and non-vanishing. We explain these observations in terms of the degeneration of some eigenvalue of the time evolution operator $U(k)$. An explicit general formula is derived for the limiting probability, from which we deduce the limiting value of the height of the observed spike at the origin. We show that the limiting probability decreases {\em quadratically} for large values of the position $x$. We locate the two minor spikes and demonstrate that their positions are determined by the phases of non-degenerated eigenvalues of $U(k)$. Finally, for fixed time $t$ sufficiently large, we examine the dependence on $t$ of the probability of finding a particle at a given location $x$.

\end{abstract}

\maketitle

\section{INTRODUCTION}

Discrete-time random walks employing a quantum coin originated in 1993 with the publication of \cite{ADZ93}. Since then, inspired by the promise of applications to the development of super-fast algorithms, especially after the publication of \cite{ NV00, AAKV01, ABNVW01} in the years 2000 and 2001, this field of research has flourished immensely.

A natural spatial setting for the evolution of a discrete-time random walk is an infinite linear lattice. In the simplest classical example, the walker starts at the origin and subsequently, depending on the outcome (heads or tails) of the toss of a fair coin, strolls one step (spatial unit) either to the right or to the left. In the quantum context, the tossing of the coin is modeled by the action on the overall state function of a ``coin operator" while the movement of the walker is modeled by the action of a ``shift operator". The observable aspects of the system are captured by the eigenvalues of these operators. The characteristically quantum-mechanical phenomenon of superposition of states gives rise to spatial probability distributions quite unlike those observed in classical random walks. Typically, sharp spikes of high probability density are observed at specific locations and troughs of near zero probability elsewhere. 

A number of cases have been studied in some depth. For instance, the  case of one particle and one coin is treated in \cite {NV00, ABNVW01, K02, K05, GJS04}; two entangled particles and one coin in \cite{OPSB06}; one partilce and many coins in \cite{BCA03, MKK07}, and one particle and two entangled coins in \cite{V-ABBB05}, etc. Very recently, Venegas-Andraca \textit{et al}. \cite{V-ABBB05} showed, by way of numerical simulations, that the walker tends to persist with high probability at the initial position as evidenced by a very prominent ``peak" or ``spike" at the origin. The occurrence of spikes and similar phenomena (called localization) also has been reported in two-dimensional Grover walks \cite{MBSS02,TFMK03,IKK04, WKKK08}, in multi-state quantum walk on a circle \cite{IK05}, in one-dimensional quantum walks driven by many coins \cite{BCA03} and in one-dimensional three-state quantum walk \cite{IKS05}.\\
{\indent}In this paper, we concentrate on the case of a discrete-time QRW on a linear lattice with two entangled coins, as proposed by Venegas-Andraca \textit{et al}. \cite{V-ABBB05}. In this framework, abbreviated ``2cQRW", the controlling behavior of the two entangled coins is modeled by a certain $4\times 4$ matrix which is defined as the second-order tensor-power of a $2\times 2$ unitary matrix. As in \cite{IKS05}, we show that the occurrence of localized spikes at the origin and elsewhere reflects the degeneracy of some eigenvalue of the time evolution operator $U(k)$. An explicit formula for the limiting spatial probability density enables us to specify the height of the spike at the origin. For sufficiently large values of the position $x$, we find that the limiting probability density decreases quadratically. This is in sharp contrast to the result obtained by Inui \textit{et al} for a three-state QRW \cite{IKS05}, where it was found that the limiting spatial probability density decreases exponentially for large $x$. 

\section{ Formulation of Quantum random walks with entangled coins on the line}

\subsection{Quantum random walks on the line with two entangled coins}

We proceed to define the elements, as formulated in \cite{V-ABBB05}, of a random walk on a discrete linear lattice governed by a pair of entangled qubits. The definitions and corresponding notations are analogous to those for the single-coin (one-qubit) framework outlined in \cite{L08}. For brevity of exposition, as in \cite{V-ABBB05}, the two-coin, or more precisely, {\em two-qubit} framework is abbreviated by the name ``2cQRW".

Akin to the one-qubit case, the coin space of the 2cQRW framework is the Hilbert space $\mathcal{H}_{ec}$ (``$ec$" for entangled coin) spanned by the orthonormal basis $\{|j\rangle;j\in B_c\}$ where $B_c=\{00, 01, 10, 11\}$. The position space is the Hilbert space $\mathcal{H}_p$ spanned by the orthonormal basis $\{|x\rangle;x \in \mathbb{Z} \}.$ The ``overall" state space of the system is $\mathcal{H}=  \mathcal{H}_{p} \otimes \mathcal{H}_{ec}$, in terms of which a general state of the system may be expressed by the formula:

\[\psi=\sum_{x\in\mathbb{Z}} \sum_{j\in B_c}\psi(x,j)|x\rangle\otimes|j\rangle.\]

The spatio-temporal progression of the 2cQRW is governed by an ``evolution operator" $U$, which is composed of a {\em coin operator} $A_{ec}$ and a {\em shift operator} $S$.

In the 2cQRW context, the coin operator is defined as the tensor product of two single-qubit operators 
$$A_{ec}=A \otimes A,$$
where the unitary operator $A$ acts on a single-coin space (see \cite{L08}). 




As in \cite{V-ABBB05}, the shift operator is given by:

\begin{eqnarray}
 S=|00\rangle\langle 00|\otimes \sum_i|i+1\rangle \langle i|+|01\rangle \langle 01|\otimes \sum_i|i\rangle \langle i| \nonumber \\
+ |10\rangle\langle 10|\otimes \sum_i|i\rangle \langle i|+|11\rangle \langle 11|\otimes \sum_i|i-1\rangle \langle i|. \label{equso}
\end{eqnarray}

The journey of the particle (a.k.a. walker) along the line is driven by a stochastic sequence of iterations of $S$. At every time step of the walk, depending on the state of the coin, the walker strolls either one spatial unit to the right or to the left or stalls at that step. Note that the walker moves only when both coins reside in the $|00\rangle$ or $|11\rangle$ state. Otherwise, the walker stalls at that step.
 
Let $I$ denote the identity operator on $\mathcal{H}_p$. Then, in terms of $A_{ec}$ and $S$, the total evolution operator $U$ is given by 
$$U = S(I\otimes A_{ec}).$$

Given $\psi_0 \in \mathcal{H}$, where $||\psi_{0}||=1$, the expression $\psi_t= U^t \psi_0$ is called the wave function for the particle at time $t$. The corresponding random walk with initial state $\psi_0$ is represented by the sequence $\{ \psi_t \}_0 ^\infty$.  

Let $X$ denote the position operator on $\mathcal{H}_p$, defined by 
$X|x\rangle=x|x\rangle$ and let $\psi_t =\sum_{x \in \mathbb{Z}}\sum_{j\in B_{c}}\psi_{t}(x, j)|x\rangle\otimes |j\rangle$ be the wave function for the particle at time $t$. Then the probability $p_t(x)$ of finding the particle at the position $x$ at time $t$ is given by the standard formula 
$$p_t(x)=\sum_{j\in B_c}|\psi_t (x, j)|^2,$$
where $|\cdot|$ indicates the modulus of a complex number.
At each instant $t$, the eigenvalues of the operator $X_t\doteq {U^{\dagger}}^{t}XU^t$ equate to the possible values of the particle's position with corresponding probability $p_t(x)$.

\subsection{Fourier transform formulation of the wave function for 2cQRW }

As in the single-coin setting treated in \cite{GJS04}, Fourier transform methods can be applied in the 2cQRW setting to obtain a useful formulation of the wave function. Let $\Psi_{t}^{ec}(x)\equiv [\psi_t (x, 1),\psi_t (x, 2),\psi_t (x, 3),\psi_t (x, 4)]^T$ represent the amplitude of the wave function, whose four components at position $x$ and time $t$ correspond respectively to the coin states 00, 01, 10, and 11. As usual, the superscript $T$ denotes the transpose operator. Assuming that the 2cQRW is launched from the origin, then the initial quantum state of the system is reflected by the components of $\Psi_{0}^{ec}(0)=[\psi_0(0,1),\psi_0(0,2),\psi_0(0,3), \psi_0(0,4)]^T\equiv [\alpha_1, \alpha_2,\alpha_3, \alpha_4]^T$, where $\sum_{j=1}^4|\alpha_j|^2=1.$

To begin, the spatial Fourier transform of $\Psi_{t}^{ec}(x)$ is defined by
$$ \widehat{\Psi_t^{ec}}(k)=\sum_{x\in \mathbb{Z}}\Psi_{t}^{ec}(x)e^{ikx}.$$
For instance, under this transformation, the initial amplitude is related to its Fourier dual by the formula:
\begin{eqnarray}
\widehat{\Psi_{0}^{ec}}(k)=\Psi_{0}^{ec}(0).
\end{eqnarray}

In general, the Fourier dual of the overall state space of the 2cQRW system is the Hilbert space $L^2(\mathbb{K})\otimes \mathcal{H}_{ec}$, consisting of $\mathbb{C}^4$-valued functions:
\begin{equation}
 \phi(k) =\left[\begin{array}{c}
  \phi_1(k)\\
   \phi_2(k) \\
    \phi_3(k) \\
     \phi_4(k)
  \end{array}\right],
\end{equation}
subject to the finiteness condition $$\Arrowvert \phi \Arrowvert^2=\Arrowvert \phi_1 \Arrowvert^2_{L^2}+\Arrowvert \phi_2 \Arrowvert^2_{L^2}+\Arrowvert \phi_3 \Arrowvert^2_{L^2}+\Arrowvert \phi_4 \Arrowvert^2_{L^2}<\infty.$$ 
Thus, given the initial state $\widehat{\Psi_{0}^{ec}}(k)$, the Fourier dual of the wave function of the 2cQRW system is expressed by
\begin{equation}
\widehat{\Psi_t^{ec}}(k)=U_{ec}(k)^t\widehat{\Psi_{0}^{ec}}(k), \label{eqnwvec}
\end{equation}
where the total evolution operator $U_{ec}(k)$ on $L^2(\mathbb{K})\otimes \mathcal{H}_{ec}$ is given by 
\begin{equation}
U_{ec}(k)= \left[\begin{array}{cccc}
e^{i k} & 0 & 0 & 0\\
0       & 1 & 0 & 0\\
0       & 0 & 1 & 0\\
0       & 0 & 0 & e^{-i k}
\end{array}\right] A_{ec}\,.\label{eqnU_{ec}}
\end{equation}

Note that $U_{ec}(k)=U(k/2)\otimes U(k/2)$, where
\begin{equation}
 U(k/2)= \left[\begin{array}{cc}
e^{i k/2} & 0\\
0 & e^{-i k/2}
\end{array}\right] A\,.\label{eqnU(k)}
\end{equation}

To wrap up this introductory section, we collect some basic facts about the eigenvalues and eigenvectors of the operators discussed above. 

Since the matrix $A$ is unitary, we may assume, without loss of generality, that its determinant is $|A|=e^{i\theta}$, where $\theta$ is a real constant. Similarly, for the unitary matrix $U(k/2)$, we may assume that its eigenvalues are $\lambda_{1}(k)=e^{i\eta(k)}$ and $\lambda_{2}(k)=e^{i(\theta-\eta(k))}$, where $\eta$ is a real-valued differentiable function of $k$. Let $v_1(k)$ and $v_2(k)$ denote the corresponding unit eigenvectors. Since $U_{ec}(k)=U(k/2)\otimes U(k/2)$, it follows that the eigenvalues of $U_{ec}(k)$ are:\vspace{.1in}\\
\hspace*{.3in}
$\left\{
\begin{array}{l}
\Lambda_1(k)=[\lambda_1(k)]^2=e^{i\varphi(k)}$ where $\varphi(k)=2\eta(k)\\
\Lambda_2(k)=\lambda_1 (k)\lambda_2(k)=|A|=e^{i\theta}\\
\Lambda_3(k)=\lambda_1 (k)\lambda_2(k)=|A|=e^{i\theta}\\
\Lambda_4(k)=[\lambda_2(k)]^2=e^{i(2\theta-\varphi(k))}.
\end{array}\right.$

Correspondingly, the unit eigenvectors are:\vspace{.1in}\\
\hspace*{.3in}$\left\{
\begin{array}{l}
V_1(k)=v_1(k)\otimes v_{1}(k)\\ 
V_2(k)=v_1(k)\otimes v_{2}(k)\\ 
V_3(k)=v_2(k)\otimes v_{1}(k)\\ 
V_4(k)=v_2(k)\otimes v_{2}(k). 
\end{array}\right.$

Finally, in terms of eigenvalues and eigenvectors, the wave function $ \widehat{\Psi_t^{ec}}(k)$ may be expanded as follows:
\begin{eqnarray}
\widehat{\Psi_t^{ec}}(k)&=& U_{ec}^t(k)\widehat{\Psi_0^{ec}}(k)\nonumber\\
\mbox{}&=& e^{it\varphi(k)}\langle V_1(k),\widehat{\Psi_0^{ec}}(k)\rangle V_1(k)\nonumber\\
\mbox{}&\mbox{}&+e^{it\theta}\langle V_2(k),\widehat{\Psi_0^{ec}}(k)\rangle V_2(k)\nonumber\\
\mbox{}&\mbox{}&+e^{it\theta}\langle V_3(k),\widehat{\Psi_0^{ec}}(k)\rangle V_3(k)\nonumber\\
\mbox{}&\mbox{}&+e^{it(2\theta-\varphi(k))}\langle V_4(k),\widehat{\Psi_0^{ec}}(k) \rangle V_4(k).\label{eqnwaec}
\end{eqnarray}

\section{Spatial probability distribution of 2cQRW}

For simplicity of notation, occasionally the explicit dependency on the parameter $k$ of the quantities $\{V_{j}\}_{j=1}^4$, $\widehat{\Psi_t^{ec}}$ and $\widehat{\Psi_0^{ec}}$ will be suppressed. By inverse Fourier transformation, the amplitude of the wave function of the particle at the position $x$ and the time $t$ is given by 
\begin{eqnarray}
\Psi_t^{ec}(x) &=& [\psi_t(x,1),\psi_t(x,2),\psi_t(x,3), \psi_t(x,4)]^T \nonumber \\
\mbox{} &=& \int_0^{2\pi}e^{-ixk}\widehat{\Psi_t^{ec}}\frac{dk}{2\pi}\nonumber\\
\mbox{} &=& I_1+I_2+I_3+I_4, \quad \label{eqnprodis1}
\end{eqnarray}

where\\

\hspace*{.3in}$\left\{
\begin{array}{l}
I_1 = \int_0^{2\pi}e^{-ixk}e^{it\varphi(k)}\langle V_1,\widehat{\Psi_0^{ec}}\rangle V_1\,\frac{dk}{2\pi}\vspace{5pt}\\ 
I_2 = \int_0^{2\pi}e^{-ixk}e^{it\theta}\langle V_2,\widehat{\Psi_0^{ec}}\rangle V_2 \,\frac{dk}{2\pi}\vspace{5pt}\\ 
I_3 = \int_0^{2\pi}e^{-ixk}e^{it\theta}\langle V_3,\widehat{\Psi_0^{ec}}\rangle V_3\,\frac{dk}{2\pi}\vspace{5pt}\\ 
I_4 = \int_0^{2\pi}e^{-ixk}e^{it(2\theta-\varphi(k))}\langle V_4,\widehat{\Psi_0^{ec}}\rangle V_4\,\frac{dk}{2\pi}\vspace{5pt}. 
\end{array}\right.$\\

In view of \cite{Eedelyi56}, the asymptotic behavior of the four Fourier integrals representing the amplitude components $I_{1}, I_{2}, I_{3} \,\,\mbox{and}\,\, I_{4}$ of $\Psi_t^{ec}(x)$ can be described by the following lemma:\vspace{5pt}

{Lemma 1.}\,\,\, Let $g(k)$ be an $N$-fold continuously differentiable function in the interval $0\leq k \leq 2\pi$. Let $g^{(n)}=d^ng/dk^n$, $A_N(x)=\sum_{n=0}^{N-1}i^{n-1}g^{(n)}(a)(-x)^{-n-1}$ and $B_N(x)=\sum_{n=0}^{N-1}i^{n-1}g^{(n)}(b)(-x)^{-n-1}$. Then, as $|x|\rightarrow \infty$, we have  
\begin{eqnarray}
\int_{0}^{2\pi}e^{-ixk}g(k)dk=B_N(x)-A_N(x)+o(x^{-N}). \label{fourierinte1}
\end{eqnarray}

The following theorem justifies the prominent spike at the origin observed by Venegas-Andraca \textit{et al} in \cite{V-ABBB05}, as well as other interesting phenomena.\vspace{5pt}  

{Theorem 1.} Let $p_t(x)\!=\!\|\Psi_t^{ec}(x)\|^2\!=\!\sum_{j\in B_c}|\psi_t (x, j)|^2$, where $\|\centerdot\|$ denotes vector norm. Let $p(x)=\lim_{t\rightarrow \infty} p_t(x)$. If the 2cQRW is launched from the origin with initial state $\widehat{\Psi_0^{ec}}$ and governed by evolution operator $U_{ec}(k)$, then the limiting probability of finding the walker at $|x\rangle$ is $$p(x)=\|\int_0^{2\pi}\{e^{-ixk}\sum_{j=2}^{3}\langle V_j(k),\widehat{\Psi_0^{ec}}(k)\rangle V_j(k)\}\frac{dk}{2\pi}\|^2.$$
 
In particular
\begin{enumerate}

\item[(i)] \, when $x=0$, we have:
\begin{eqnarray}
p(0)=\|\int_0^{2\pi}\{\sum_{j=2}^{3}\langle V_j(k),\widehat{\Psi_0^{ec}}(k)\rangle V_j(k)\}\frac{dk}{2\pi}\|^2. \label{spike-0}
\end{eqnarray}

\item [(ii)]\,as $|x|\rightarrow \infty$, we have: 
\begin{eqnarray}
p(x)&\sim & \frac{1}{x^2}\|\sum_{j=2}^3[\langle V_j(0), \widehat{\Psi_0^{ec}}(0)\rangle V_j(0)- \nonumber \\
\mbox{}&\mbox{}&\langle V_j(2\pi), \widehat{\Psi_0^{ec}}(2\pi)\rangle V_j(2\pi)]\|^2 
+O(x^{-2}). \label{prob-x}
\end{eqnarray}
\end{enumerate}

Proof. See Appendix A.\vspace{5pt}

The height of the spike at the origin is quantified precisely by Eq.(\ref{spike-0}).  For instance, if we choose, as the initial coin state, the Bell state $|\Phi^{+}\rangle=\frac{1}{\sqrt{2}}\left(|00\rangle +|11\rangle\right)$, then the probability of finding the particle at the origin converges to $p(0)= 3-2\sqrt{2}\thickapprox 0.171573$ (justification below). This agrees quite accurately with the graphical representation in FIG. 1 (see below) and the observations reported in \cite{V-ABBB05}. No such spikes are evident in QRWs driven by single coins.

The formula Eq.(\ref{prob-x}) for $p(x)$ implies that the limiting probability of finding the particle at any fixed location \underline{does} \underline{not} vanish. This is in sharp contrast to the behavior of single-coin QRWs \cite{NV00, ABNVW01, K02, K05}, for which the analogous limiting probability converges everywhere to zero. 

In \cite{IKS05}, which treats the case of a one-dimensional three-state quantum walk, there too the  behavior of $p(x)$ is observed to be everywhere non-vanishing. However, one important difference distinguishes that case from the present case.  Whereas in \cite{IKS05} the rate of decay of $p(x)$ as $|x|\rightarrow \infty$ is {\em exponential}, in the present case the rate of decay of $p(x)$  as $|x|\rightarrow \infty$ is {\em quadratic}.

By the proof of Theorem 1 (see Appendix A), the everywhere non-vanishing behavior of $p(x)$ as well as the occurrence of the spike at the origin both are due to the independence on $k$ of the degenerate eigenvalues of the evolution operator $U_{ec}$. 

We remark that the sequence $\{p(x)\}_{x=0}^{\infty}$ of limiting probabilities does not itself amount to a probability distribution. In fact the sum $\sum_{x\in \mathbb{Z}}p(x)$ usually is less than unity. The following theorem evaluates this sum.\vspace{5pt}

{Theorem 2.}\,\, If the 2cQRW is launched from the origin with initial state $\widehat{\Psi_0^{ec}}$ and governed by evolution operator $U_{ec}(k)$, then 
\begin{eqnarray}
\sum_{x\in \mathbb{Z}}p(x)
=\int_{0}^{2\pi}\sum_{j=2}^3|\langle V_j(k),\widehat{\Psi_0^{ec}}(k)\rangle|^2 \frac{dk}{2\pi}. \label{localization}
\end{eqnarray}
Proof. See Appendix B.\vspace{5pt}

To illustrate the above theorem, suppose the 2cQRW is launched from the origin with initial state $|\Phi^{+}\rangle=\frac{1}{\sqrt{2}}\left(|00\rangle +|11\rangle\right)$, governed by the coin operator $H^{\otimes 2}$, where $H$ is the $2\times 2$ Hadamard operator. Then, by (\ref{localization}), $\sum_{x\in \mathbb{Z}}p(x)=\sqrt{2}-1$. This example is discussed further in the sequel.

The third and final theorem of this article provides detailed information on the local behavior of $p_t(x)$, the probability of finding the particle at time $t$ and position $x$. In preparation for this theorem, we digress for a moment to review some prerequisites.

As in Eq.(\ref{eqnU_{ec}}), let $\varphi(k)$ denote the {\em phase} of the principal eigenvalue $\Lambda_1=e^{i\varphi(k)}$ associated with the evolution operator $U_{ec}(k)$. To ensure sufficient control over the analytic properties of $\varphi$, our choice of the coin operator $A_{ec}$ must be restricted accordingly. Specifically, $\varphi$ must possesses no {\em stationary points} (see \cite{NV00, Eedelyi56}) of order greater than 1. Moreover, the maximum and minimum values of the function $\varphi^{\prime}(k)$ over $[0,2\pi]$ must coincide respectively with $|\varphi^{\prime}(k_0)|$ and $-|\varphi^{\prime}(k_0)|$, where $k_0$ is a zero of $\varphi^{\prime\prime}(k)$ of multiplicity one. 

Provided that $\beta$ is neither 0 nor $\frac{\pi}{2}$, the aforementioned assumptions are valid at least for the important case of the coin operator $A(\beta)^{\otimes 2}$ treated in \cite{L08}. If $\beta$ is 0 or $\frac{\pi}{2}$, then the resulting QRW is trivial. 

In the special case when $\beta=\frac{\pi}{4}$, our coin operator $A(\frac{\pi}{4})^{\otimes 2}$ coincides with the $4\times 4$ Hadamard-based matrix $H^{\otimes 2}$. This example is fully  elaborated below.\vspace{5pt} 

{Theorem 3.}\,\,\, Suppose the 2cQRW is launched from the origin with initial state $\widehat{\Psi_0^{ec}}(k)$ and governed by evolution operator $U_{ec}(k)$ based on the coin operator $A_{ec}=A(\beta)^{\otimes 2}$. Let $\varphi(k)$ denote the phase function  of the principal non-degenerate eigenvalue $\Lambda_1=e^{i\varphi(k)}$ of $U_{ec}(k)$ and let $M=|\varphi^{\prime}(k_0)|$, where $k_0$ is a zero of $\varphi^{\prime\prime}(k)$.
 
\begin{enumerate}
\item[(i)] When $x=0$, we have
 \begin{eqnarray}
p_t(0)&=& p(0)+O(t^{-\frac{1}{2}}) \label{eqnprodis}.
\end{eqnarray} 

\item[(ii)] If $x$ is an integer within a small fixed distance $\delta$ of $\pm{tM}$ then $p_t(x)=O(t^{-\frac{2}{3}})$.

 \item[(iii)] For a fixed $\epsilon > 0$, if $t\geq {x} \geq t(M+\epsilon)$ or $t\geq -{x} \geq t(M+\epsilon)$, then $p_t(x)=O(t^{-2})$.

\item[(iv)] For a fixed $\epsilon > 0$, if $t^{\frac{1}{2}}\leq x \leq t(M-\epsilon)$ or $t^{\frac{1}{2}}\leq -x \leq t(M-\epsilon)$, then $p_t(x)=O(t^{-1})$.

 \item[(v)] If $|x|<t^{\frac{1}{2}}$, and $|x|\approx 0$, then 
\begin{eqnarray}
p_t(x)&=&p(x)+O(t^{-\frac{1}{2}})\label{eqnprodis}.
\end{eqnarray}

\item[(vi)] If $|x|<t^{\frac{1}{2}}$, and $|x|\approx t^{\frac{1}{2}}$, then $p_t(x)=O(t^{-1}).$ 

\end{enumerate}\vspace{5pt}

According to this theorem, if $t$ is fixed and sufficiently large, the distribution of $p_t(x)$ is characterized by a spike at the origin and two minor spikes located at $\pm t\varphi^{\prime}(k_{0})$. As $t$ increases, the two minor spikes shrink in height and drift off to infinity in both directions (see FIG. 1), while the spike at the origin persists and settles to a specific height given by Eq.(\ref{spike-0}). These inferences confirm the observations based on numerical simulations reported in \cite{V-ABBB05}.\vspace{5pt}

{Example.}\,\,\, For the duration of this example, let us agree to adopt the convection whereby any row vector, when enclosed in round parentheses, is identified with the corresponding column vector enclosed in square brackets. For instance:
\[(a_{1},a_{2},a_{3})=\left[\begin{array}{c}a_{1}\\a_{2}\\a_{3}\end{array}\right]\]

Suppose the 2cQRW is launched from the origin with initial state $|\Phi^{+}\rangle=\frac{1}{\sqrt{2}}\left(|00\rangle +|11\rangle\right)$ and governed by the coin operator 
$$H^{\otimes 2}=\left[\frac{1}{\sqrt{2}}\left(|0\rangle \langle 0|+|0\rangle \langle 1|+|1\rangle \langle 0|-|1\rangle \langle 1|\right)\right]^{\otimes 2},$$
which equals the operator $A_{ec}=A(\beta)^{\otimes 2}$ in Theorem 3 when $\beta=\frac{\pi}{4}$.

The total evolution operator is given by $U_{ec}(k)=U(k/2) \otimes U(k/2)$, where 
\begin{equation}
 U(k/2)= \frac{1}{\sqrt{2}}\left[\begin{array}{cc}
e^{i k/2} & e^{i k/2}\\
e^{-i k/2} &-e^{-i k/2}
\end{array}\right]. \label{eqnU(k/2)}
\end{equation}

Let $\varphi(k)=2\sin^{-1}\left(\frac{\sin(k/2)}{\sqrt{2}}\right)$. Then the two eigenvalues of $U(k/2)$ are
\[\begin{array}{l}
\lambda_1(k)=e^{i\varphi(k)/2}\vspace{7pt}\\ 
\lambda_2(k)=-e^{-i\varphi(k)/2}. 
\end{array}\]\vspace{5pt}
and the four eigenvalues of $U_{ec}(k)=U(k/2) \otimes U(k/2)$ are:\vspace{5pt}
\hspace*{.3in}
$\left\{
\begin{array}{l}
\Lambda_1(k)=\lambda_1(k)^2=e^{i\varphi(k)}\\
\Lambda_2(k)=\lambda_1 (k)\lambda_2(k)=-1\\
\Lambda_3(k)=\lambda_1 (k)\lambda_2(k)=-1\\
\Lambda_4(k)=\lambda_2(k)^2=e^{-i\varphi(k)}.
\end{array}\right.$\vspace{5pt}

To apply Theorem 3, observe that  
\begin{eqnarray*}
\varphi^{\prime}(k)=\frac{\cos(k/2)}{\sqrt{2-\sin^2(k/2)}},\\
\varphi^{\prime \prime}(k)=\frac{-\sin(k/2)}{2[2-\sin^2(k/2)]^{\frac{3}{2}}}.
\end{eqnarray*} 
so that the zeros of $\varphi^{\prime \prime}(k)$, both of multiplicity one, occur at $0$ and $2\pi$. Thus, the maximum value $M=\frac{\sqrt{2}}{2}$ of $\varphi^{\prime}(k)$ is attained at $k=0$ while its  minimum  value $-\frac{\sqrt{2}}{2}$ is attained at $k=2\pi$. 

For economy of notation, let 
\[\begin{array}{l}
\gamma_{1}(k)=-\cos\frac{k}{2}+\sqrt{1+\cos^2\frac{k}{2}}\vspace{7pt}\\ 
\gamma_{2}(k)=-\cos\frac{k}{2}-\sqrt{1+\cos^2\frac{k}{2}}, 
\end{array}\]
and
\[\begin{array}{l}
N_{1}(k)=2-2\gamma_{1}(k)\cos\frac{k}{2}\vspace{7pt}\\ 
N_{2}(k)=2-2\gamma_{2}(k)\cos\frac{k}{2}, 
\end{array}\]
in terms of which, the two unit eigenvectors of $U(k/2)$ corresponding to the eigenvalues $\lambda_1(k)$ and $\lambda_2(k)$ are:
\[\begin{array}{l}
v_{1}(k)=\frac{1}{\sqrt{N_1}}\left(e^{ik/2},\gamma_{1}(k)\right)\vspace{5pt}\\
v_{2}(k)=\frac{1}{\sqrt{N_2}}\left(e^{ik/2},\gamma_{2}(k)\right), 
\end{array}\]
and the four unit eigenvectors of $U_{ec}(k)$ are:\vspace{5pt}\\
\hspace*{10pt}$\left\{
\begin{array}{l}
V_1(k)=\frac{1}{N_1}\left(e^{ik}, e^{ik/2}\gamma_{1}(k), e^{ik/2}\gamma_{1}(k), \gamma_{1}(k)^{2}\right)\vspace{5pt}\\ 
V_2(k)=\frac{1}{\sqrt{N_1N_2}}\left( e^{ik}, e^{ik/2}\gamma_{2}(k), e^{ik/2}\gamma_{1}(k),-1 \right)\vspace{5pt}\\ 
V_3(k)=\frac{1}{\sqrt{N_1N_2}}\left(e^{ik}, e^{ik/2}\gamma_{1}(k), e^{ik/2}\gamma_{2}(k), -1 \right)\vspace{5pt}\\ 
V_4(k)=\frac{1}{N_2}\left(e^{ik}, e^{ik/2}\gamma_{2}(k), e^{ik/2}\gamma_{2}(k), \gamma_{2}(k)^{2}\right)\vspace{5pt}. 
\end{array}\right.$

By Eq.(\ref{eqnwaec}), with $\widehat{\Psi _0^{ec}}(k)=\left(\frac{\sqrt{2}}{2},0,0,\frac{\sqrt{2}}{2}\right)$, the wave function $\widehat{\Psi_t^{ec}}(k)=U_{ec}^t(k)\widehat{\Psi_0^{ec}}(k)$ may be expressed as a sum of four components 
\begin{equation}\widehat{\Psi_t^{ec}}(k)=H_{1}(k,t)+H_{2}(k,t)+H_{3}(k,t)+H_{4}(k,t),\nonumber\end{equation}
where:\vspace{5pt}\\
\hspace*{10pt}$\left\{
\begin{array}{l}
H_{1}(k,t)=\frac{\sqrt{2}e^{it\varphi(k)}}{2N_1}\left(e^{-ik}+\gamma_{1}(k)^{2}\right)V_{1}(k)\vspace{5pt}\\ 
H_{2}(k,t)=\frac{\sqrt{2}e^{i\pi t}}{2\sqrt{N_1N_2}}\left( e^{-ik}-1 \right)V_{2}(k)\vspace{5pt}\\ 
H_{3}(k,t)=\frac{\sqrt{2}e^{i\pi t}}{2\sqrt{N_1N_2}}\left(e^{-ik} -1 \right)V_{3}(k)\vspace{5pt}\\ 
H_{4}(k,t)=\frac{\sqrt{2}e^{it\varphi(k)}}{2N_2}\left(e^{-ik}+\gamma_{2}(k)^{2}\right)V_{4}(k).\vspace{5pt} 
\end{array}\right.$

Applying inverse Fourier transformation, as in Eq.(\ref{eqnprodis1}), the four summands of 
\begin{equation}\Psi_t^{ec}(x)=I_{1}+I_{2}+I_{3}+I_{4}\label{4Is}\end{equation} 
are:\vspace{5pt}

\hspace*{3pt}$\left\{
\begin{array}{l}
I_1 = \int_0^{2\pi}e^{-ixk}\frac{\sqrt{2}e^{it\varphi(k)}}{2N_1}\left(e^{-ik}+\gamma_{1}(k)^{2}\right)V_{1}(k)\,\frac{dk}{2\pi}\vspace{5pt}\\ 
I_2 = \int_0^{2\pi}e^{-ixk}\frac{\sqrt{2}e^{i\pi t}}{2\sqrt{N_1N_2}}\left( e^{-ik}-1 \right)V_{2}(k)\vspace{5pt}\,\frac{dk}{2\pi}\vspace{5pt}\\ 
I_3 = \int_0^{2\pi}e^{-ixk}\frac{\sqrt{2}e^{i\pi t}}{2\sqrt{N_1N_2}}\left(e^{-ik} -1 \right)V_{3}(k)\vspace{5pt}\,\frac{dk}{2\pi}\vspace{5pt}\\ 
I_4 = \int_0^{2\pi}e^{-ixk}\frac{\sqrt{2}e^{it\varphi(k)}}{2N_2}\left(e^{-ik}+\gamma_{2}(k)^{2}\right)V_{4}(k)\,\frac{dk}{2\pi}\vspace{5pt}. 
\end{array}\right.$\\

To determine the height of the spike at the origin, as given by Eq.(\ref{spike-0}), we must evaluate the sum of the second and third terms of 
(\ref{4Is}) at $x=0$. After some algebraic manipulations, using the identity $N_{1}N_{2}=4+4\cos^{2}\frac{k}{2}$ and the formulas

\begin{eqnarray*}
\int_0^{2\pi}\frac{1}{1+\cos^2\frac{k}{2}}\frac{dk}{2\pi}&=&\frac{\sqrt{2}}{2},\\
\int_0^{2\pi}\frac{\cos k}{1+\cos^2\frac{k}{2}}\frac{dk}{2\pi}&=&2-\frac{3\sqrt{2}}{2},\\
\int_0^{2\pi}\frac{\sin k}{1+\cos^2\frac{k}{2}}\frac{dk}{2\pi}&=&0,
\end{eqnarray*}
we arrive at the formula:
\begin{equation}I_{2}+I_{3}=\frac{1}{2}\left((-1)^t(2-\sqrt{2}),0,0, (-1)^t(2-\sqrt{2})\right).\label{eqn23}\end{equation}

Therefore, the probability of finding the particle at the origin converges to 

\begin{eqnarray*}
p(0)&=&\|I_{2}+I_{3}\|\\
&=&3-2\sqrt{2}\\
&\thickapprox& 0.171573.
\end{eqnarray*}

Our computer simulation (see FIG. 1) accords well with this theoretical value as does the simulation result 0.171242 reported in \cite{V-ABBB05}.

\begin{figure}
\includegraphics[height=2.0in]{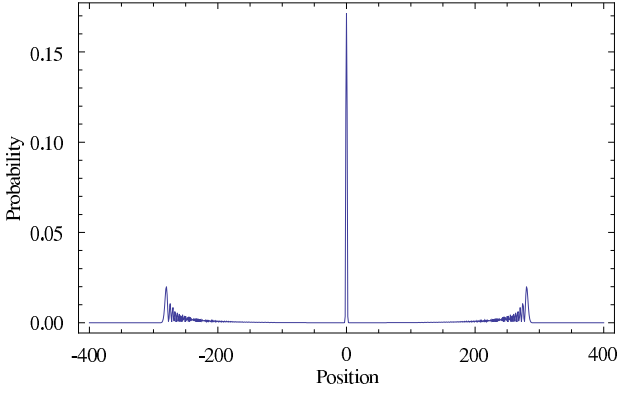}
\caption{\label{fig:wide}The position probability distribution $p_{t}(x)$ for a 2cQRW with initial coin state $|\Phi^{+}\rangle=\frac{1}{\sqrt{2}}\left(|00\rangle +|11\rangle\right)$ after $t=400$ steps.}
\end{figure}

We proceed to elucidate the destinies of the two minor spikes, which plainly are visible in FIG. 1. By Theorem 3, with $M=\frac{\sqrt{2}}{2}$, each of the spikes, respectively, is predicted to lurk within a small neighborhood of the positions $x=\pm \frac{\sqrt{2}}{2}t$. It suffices to consider only the spike at $x=\frac{\sqrt{2}}{2}t$, since the other one may be treated similarly. Also, to serve as our small neighborhood, it suffices to adopt the interval $J_{t}=[\frac{\sqrt{2}}{2}t-1,\frac{\sqrt{2}}{2}t+1]$. 

Let $x=\frac{\sqrt{2}}{2}t+c\in J_{t}$. Substituting in each of the summands of Eq.(\ref{4Is}) this value of $x$ and performing some simple algebraic manipulations, we obtain:\vspace{5pt}

{\hspace*{-10pt}}
$\left\{
\begin{array}{l}
I_1 = \int_0^{2\pi}\!\!e^{it[\varphi(k)-\frac{\sqrt{2}}{2}k]}e^{-ik\delta}\frac{\sqrt{2}}{2N_1} \left(e^{-ik}+\gamma_{1}(k)^{2}\right)V_{1}(k)\,\frac{dk}{2\pi}\vspace{5pt}\\ 
I_2 = \int_0^{2\pi}e^{i\pi t}e^{-ixk}\frac{\sqrt{2}}{2\sqrt{N_1N_2}} \left( e^{-ik}-1 \right)V_{2}(k)\vspace{5pt}\,\frac{dk}{2\pi}\vspace{5pt}\\ 
I_3 = \int_0^{2\pi}e^{i\pi t}e^{-ixk}\frac{\sqrt{2}}{2\sqrt{N_1N_2}} \left(e^{-ik} -1 \right)V_{3}(k)\vspace{5pt}\,\frac{dk}{2\pi}\vspace{5pt}\\ 
I_4 = \int_0^{2\pi}\!\!e^{it[-\varphi(k)-\frac{\sqrt{2}}{2}k]}e^{-ik\delta}\frac{\sqrt{2}}{2N_2}\!\left(e^{-ik}\!+\!\gamma_{2}(k)^{2}\right)\!V_{4}(k)\,\frac{dk}{2\pi}\vspace{5pt}. 
\end{array}\right.$

By Lemma 1, both $I_2$ and $I_3$ decay as $O(t^{-1})$. Hence, as $t\rightarrow\infty$, their contributions to the value of $p_{t}(x)$ become eventually negligible relative to $I_{1}$ and $I_{4}$, whose values, as shown below, decay no faster than $O(t^{-\frac{2}{3}})$.
  
Without much effort, one verifies that the modified phase function $\varphi(k)-\frac{\sqrt{2}}{2}k$ possesses a stationary point of order 2 at $k=0$ and, by similar reasoning, $-\varphi(k)-\frac{\sqrt{2}}{2}k$ possesses a stationary point of order 2 at $k=2\pi$. Thus, by the method of ``stationary phase" \cite{NV00, Eedelyi56}, the following values are obtained for the leading terms of $I_1$ and $I_4$:\vspace{5pt}
\begin{eqnarray*}
I_1 &\thicksim& C_{1}\cdot (1,\sqrt{2}-1,\sqrt{2}-1,3-2\sqrt{2}),\\
I_4 &\thicksim& C_{2}\cdot (1,\sqrt{2}-1,\sqrt{2}-1,3-2\sqrt{2}),
\end{eqnarray*}where
\begin{eqnarray*}
C_{1}&=&\frac{1}{2\pi}\frac{2\sqrt{2}-2}{(4-2\sqrt{2})^2} e^{-i\frac{\pi}{6}}\left(\frac{96}{t\sqrt{2}}\right)^{\frac{1}{3}}\frac{\Gamma(\frac{1}{3})}{3}\nonumber \\
C_{2}&=&\frac{1}{2\pi}\frac{2\sqrt{2}-2}{(4-2\sqrt{2})^2}e^{-ic2\pi} e^{-it\sqrt{2}\pi+i\frac{\pi}{6}}\left(\frac{96}{t\sqrt{2}}\right)^{\frac{1}{3}}\frac{\Gamma(\frac{1}{3})}{3}.
\end{eqnarray*}

Finally, after some further manipulations, we obtain:
\begin{eqnarray*}
p_t(x)&\thicksim& \|I_1+I_4\|^2\\
&\thicksim& \|e^{-i\frac{\pi}{6}}+e^{-ic2\pi}e^{-it\sqrt{2}\pi+i\frac{\pi}{6}}\|^2
\cdot \frac{(6\sqrt{2})^{\frac{2}{3}}\Gamma(\frac{1}{3})^2}{18\pi^2}t^{-\frac{2}{3}}\\
&\thicksim&\frac{(6\sqrt{2})^{\frac{2}{3}}\Gamma(\frac{1}{3})^2}{6\pi^2}t^{-\frac{2}{3}}.
\end{eqnarray*}

\section{RELATED IDEAS FOR FURTHER RESEARCH}

As above, let $X$ denote the position operator on the position space $\mathcal{H}_p$ and let $X_t\doteq {U^{\dagger}}^{t}XU^t$, where $U$ denotes the evolution operator. A natural and potentially more useful statistical description of the evolution of the 2cQRW as time $t\rightarrow \infty$ can be developed in terms of the ``normalized" operator $\frac{1}{t}X_{t}$. 

Suppose the 2cQRW, controlled by the coin operator $H^{\otimes 2}$, is launched from the origin in the initial state $\Psi_{0}^{ec}(0)=\alpha_1|00\rangle+\alpha_2|01\rangle+\alpha_3|10\rangle+\alpha_4|11\rangle$, where $\sum_{j=1}^4|\alpha_j|^2=1$. For $y\in [-1,1]$, let $\delta_0(y)$ denote the \textit{point mass at the origin} and let $I_{(a,b)}(y)$ denote the \textit{indicator function} of the real interval $(a,b)$. Then, as $t\rightarrow \infty$, the normalized position distribution $f_{t}(y)$ associated with $\frac{1}{t}X_{t}$ converges, in the sense of a weak limit, to the density function

\begin{eqnarray}
f(y)=c_{00}\delta_0(y)+\frac{I_{(-1/\sqrt{2},1/\sqrt{2})}(y)}{\pi(1-y^2)\sqrt{1-2y^2}}\sum_{j=0}^2c_j y^j.\label{generaldensityfn}
\end{eqnarray}

In the above formula, the coefficients $c_{00}$, $c_{0}$, $c_{1}$ and $c_{2}$ are given by 
\begin{eqnarray}
c_{00}&=&\frac{\sqrt{2}}{4}+\frac{1}{2}(2-\sqrt{2})(|\alpha_2|^2+|\alpha_3|^2)\nonumber\\
\mbox{}&\mbox{}&+\frac{1}{2}\mathrm{Re}[(2-\sqrt{2})(\alpha_2\overline{\alpha_4}+\alpha_3\overline{\alpha_4}-\alpha_1\overline{\alpha_2}-\alpha_1\overline{\alpha_3})\nonumber\\
\mbox{}&\mbox{}&+(3\sqrt{2}-4)\alpha_1 \overline{\alpha_4}-\sqrt{2}\alpha_2\overline{\alpha_3}]\label{c_{00}}\\
c_{0}&=&\frac{1}{2}+\mathrm{Re}\{\alpha_2\overline{\alpha_3}-\alpha_1\overline{\alpha_4}\}  \label{c_0}\\
c_{1}&=&|\alpha_1|^2-|\alpha_4|^2\nonumber\\
\mbox{}&\mbox{}&+\mathrm{Re}(\alpha_1\overline{\alpha_2}+\alpha_1\overline{\alpha_3}+\alpha_2\overline{\alpha_4}+\alpha_3\overline{\alpha_4})\nonumber\\
c_{2}&=&\frac{1}{2}(|\alpha_1|^2+|\alpha_4|^2-|\alpha_2|^2-|\alpha_3|^2)\nonumber\\
\mbox{}&\mbox{}&+\mathrm{Re}[3\alpha_1\overline{\alpha_4}+\alpha_1\overline{\alpha_2}+\alpha_1\overline{\alpha_3}\nonumber\\
\mbox{}&\mbox{}&-\alpha_2\overline{\alpha_3}-\alpha_2\overline{\alpha_4}-\alpha_3\overline{\alpha_4})]\nonumber.
\end{eqnarray}

The derivation of Eq.(\ref{generaldensityfn}) is due to the method by Grimmett \textit{et al.} \cite{GJS04}.

For instance, if the 2cQRW is launched from the Bell state $|\Phi^{+}\rangle=\frac{1}{\sqrt{2}}(|00\rangle +|11\rangle)$, then we obtain 

\begin{eqnarray}
f(y)=(\sqrt{2}-1)\delta_0(y)+\frac{2y^2I_{(-1/\sqrt{2},1/\sqrt{2})}(y)}{\pi(1-y^2)\sqrt{1-2y^2}}.\label{specificdensityfn}
\end{eqnarray}

It is no accident that the coefficient $c_{00}=\sqrt{2}-1$ of $\delta_{0}(y)$ in Eq. (\ref{specificdensityfn}) coincides with the value of the sum in Eq. (\ref{localization}). In this particular instance, we have  

$$c_{00}=\sum_{x\in \mathbb{Z}}p(x)=\sqrt{2}-1.$$

A detailed study of the limit distribution of $\frac{1}{t}X_t$ for 2cQRW will be presented in a forthcoming publication.

\begin{acknowledgments}
We thank our undergraduate research assistants, Jamin Gallman and Brian Cunningham, for helping with the computer simulation to generate the graph in FIG. 1.  The authors gratefully acknowledge support from Project HBCU-UP/BETTER at Bowie State University. 
\end{acknowledgments}

\appendix

\section{PROOF OF THEOREM 1}

Proof. \,\,\, Let $m\geq 0$ denote the highest order of the stationary points of $\varphi(k)$. By the method of stationary phase \cite{NV00,Eedelyi56} applied to the first term and the fourth term in Eq. (\ref{eqnprodis1}), we have

\begin{eqnarray} 
\int_0^{2\pi}\{e^{-ixk}e^{it\varphi(k)}\langle V_1(k),\widehat{\Psi_0^{ec}}(k)\rangle V_1(k)\}\frac{dk}{2\pi}
\nonumber\\
\rightarrow O(t^{-\frac{1}{m+1}})\,\,\,\, \mathrm{as\,\, t\rightarrow \infty;} \label{eqnv_1}
\end{eqnarray}
\begin{eqnarray} 
\int_0^{2\pi}\{e^{-ixk}e^{it(2\theta-\varphi(k))}\langle V_4(k),\widehat{\Psi_0^{ec}}(k)\rangle V_4(k)\}\frac{dk}{2\pi}
\nonumber\\
\rightarrow O(t^{-\frac{1}{m+1}})\,\,\,\, \mathrm{as\,\, t\rightarrow \infty.} \label{eqnv_4}
\end{eqnarray}

Since the eigenvalues corresponding to $I_2$,  $I_3$ in Eq. (\ref{eqnprodis1}) are independent of $k$, the values of $I_2$ and $I_3$ do not vanish as $t\rightarrow \infty$. Hence, $I_2$ and $I_3$ provide the only non-negligible contributions to the limiting amplitude. Thus

$$p(x)=\left\|\int_0^{2\pi}\{e^{-ixk}\sum_{j=2}^{3}\langle V_j(k),\widehat{\Psi_0^{ec}}(k)\rangle V_j(k)\}\frac{dk}{2\pi}\right\|^2.$$

By Lemma 1, as $x\rightarrow\infty$, we obtain 

\begin{eqnarray}
\int_0^{2\pi}\{e^{-ixk}\sum_{j=2}^{3}\langle V_j(k),\widehat{\Psi_0^{ec}}(k)\rangle V_j(k)\}\frac{dk}{2\pi} \nonumber \\
=\frac{1}{ix}\sum_{j=2}^3[\langle V_j(0), \widehat{\Psi_0^{ec}}(0)\rangle V_j(0)- \nonumber \\
\langle V_j(2\pi), \widehat{\Psi_0^{ec}}(2\pi)\rangle V_j(2\pi)]+O(x^{-1}) \label{eqnv_2-3}.
\end{eqnarray}
Therefore, 
\begin{eqnarray}
p(x)\sim \frac{1}{x^2}\|\sum_{j=2}^3[\langle V_j(0), \widehat{\Psi_0^{ec}}(0)\rangle V_j(0)- \nonumber \\
\langle V_j(2\pi), \widehat{\Psi_0^{ec}}(2\pi)\rangle V_j(2\pi)]\|^2 
+o(x^{-2}).
\end{eqnarray}

At the other extreme, setting $x=0$, we obtain:

 \begin{eqnarray}
p(0)=\|\int_0^{2\pi}\{\sum_{j=2}^{3}\langle V_j(k),\widehat{\Psi_0^{ec}}(k)\rangle V_j(k)\}\frac{dk}{2\pi}\|^2. 
\end{eqnarray}.

\section{PROOF OF THEOREM 2}

Proof. \,\,Let
\[W(k)=\sum_{j=2}^{3}\langle V_j(k), \widehat{\Psi_0^{ec}}(k)\rangle V_j(k),\]
and let
\[c_{x}=\int_{0}^{2\pi}e^{-ixk}W(k)\frac{dk}{2\pi},\]
so that
\[W(k)=\sum_{x\in\mathbb{Z}}c_xe^{ixk}.\]

Since
\[\langle W(k),W(k)\rangle=\sum_{x\in \mathbb{Z}}\|c_x\|^2+\sum_{x\neq y}c_x\overline{c_y}e^{i(x-y)k},\]
and, by Theorem 1, 
\[\sum_{x\in \mathbb{Z}}p(x)=\sum_{x\in \mathbb{Z}}\|c_x\|^2,\] 
we conclude that
\begin{eqnarray*}
\int_0^{2\pi}\langle W(k),W(k)\rangle\frac{dk}{2\pi}&=&\sum_{x\in \mathbb{Z}}\|c_x\|^2\\
                                             \mbox{}&=&\sum_{x\in \mathbb{Z}}p(x).
\end{eqnarray*}.

On the other hand, 
\[\langle W(k),W(k)\rangle=\sum_{j=2}^3|\langle V_j(k), \widehat{\Psi_0^{ec}}(k)\rangle|^2.\]
Therefore
\[\sum_{x\in \mathbb{Z}}p(x)=\int_0^{2\pi}\sum_{j=2}^3 |\langle V_j(k), \widehat{\Psi_0^{ec}}(k)|^2\frac{dk}{2\pi}.\]

\section{PROOF OF THEOREM 3}

Proof. \,\,(i) Assuming that 1 is the highest order of any stationary point of $\varphi(k)$, we infer, by the method of stationary phase \cite{NV00,Eedelyi56}, that the first term $I_1$ and the fourth term $I_{4}$ in Eq.(\ref{eqnprodis1}) decay as follows:

\begin{eqnarray} 
\int_0^{2\pi}\{e^{-ixk}e^{it\varphi(k)}\langle V_1(k),\widehat{\Psi_0^{ec}}(k)\rangle V_1(k)\}\frac{dk}{2\pi}
\nonumber\\
\rightarrow O(t^{-\frac{1}{2}})\,\,\,\, \mathrm{as\,\, t\rightarrow \infty;} \label{eqnv_1}
\end{eqnarray}
\begin{eqnarray} 
\int_0^{2\pi}\{e^{-ixk}e^{it(2\theta-\varphi(k))}\langle V_4(k),\widehat{\Psi_0^{ec}}(k)\rangle V_4(k)\}\frac{dk}{2\pi}
\nonumber\\
\rightarrow O(t^{-\frac{1}{2}})\,\,\,\, \mathrm{as\,\, t\rightarrow \infty.} \label{eqnv_4}
\end{eqnarray}
Hence
 \begin{eqnarray}
p_t(0)&=&\|\int_0^{2\pi}
\sum_{j=2}^3\langle V_j(k),\widehat{\Psi_0^{ec}}(k)\rangle V_j(k) \frac{dk}{2\pi}\|^2 
 +O(t^{-\frac{1}{2}})\nonumber \\
&=&p(0)+O(t^{-\frac{1}{2}}).
\end{eqnarray}

(ii)\, \,If $x\in[t\varphi^{\prime}(k_{0})-\delta,t\varphi^{\prime}(k_{0})+\delta]$, say, $x=t\varphi^{\prime}(k_0)+c$, where $|c|\leq \delta$, then the amplitudes of Eq.(\ref{eqnprodis1}) can be expressed as follows:

\begin{eqnarray}
(\psi_t(x,1),\psi_t(x,2),\psi_t(x,3), \psi_t(x,4) )^T \nonumber \\
=\int_0^{2\pi}\{e^{it[\varphi(k)-\varphi^{\prime}(k_0)k]}e^{-ic k}\langle V_1,\widehat{\Psi_0^{ec}}\rangle V_1\nonumber\\
+e^{it[\theta-k\varphi^{\prime}(k_0)]}e^{-ic k}\langle V_2,\widehat{\Psi_0^{ec}}\rangle V_2\nonumber\\
+e^{it[\theta-k\varphi^{\prime}(k_0)]}e^{-ic k}\langle V_3,\widehat{\Psi_0^{ec}}\rangle V_3\nonumber\\
+e^{it[2\theta-\varphi(k)-\varphi^{\prime}(k_0)k]}e^{-ic k}\langle V_4,\widehat{\Psi_0^{ec}} \rangle V_4\} \frac{dk}{2\pi}. \quad \label{eqnprodis2}
\end{eqnarray}

Since $k_0$ is a zero of $\varphi^{\prime\prime}(k)$ with multiplicity 1, $k_0$ is a stationary point of $\varphi(k)-\varphi^{\prime}(k_0)k$, $\theta-k\varphi^{\prime}(k_0)$ and $2\theta-\varphi(k)-\varphi^{\prime}(k_0)k$ of order 2, 0, and 0, respectively.

By the method of stationary phase, as in \cite{NV00,Eedelyi56}, we infer that, as $t\rightarrow \infty$,
the first term $I_1$ in Eq. (\ref{eqnprodis2}) is the order of $O(t^{-\frac{1}{3}})$, while the other three terms in 
Eq. (\ref{eqnprodis2}) are of order $O(t^{-1})$. Therefore, as $t\rightarrow \infty$, $\psi_t(x,j)=O(t^{-\frac{1}{3}})$ where $j=1,2,3,4$.

Similarly, when $x\in[-t\varphi^{\prime}(k_{0})-\delta,-t\varphi^{\prime}(k_{0})+\delta]$, we get $\psi_t(x,j)\sim O(t^{-\frac{1}{3}})$where $j=1,2,3,4$. Hence, in either case, $p_t(x)=O(t^{-\frac{2}{3}})$.

(iii)\,\, For convenience, we may assume $\varphi^{\prime}(k_0)>0$, then $M=\varphi^{\prime}(k_0)$. It suffices to treat the case when $t\geq x>(\varphi^{\prime}(k_0)+\epsilon)t$, as other case can be treated similarly. The amplitudes of in Eq. (\ref{eqnprodis1}) can  be expressed as follows: 

\begin{eqnarray}
(\psi_t(x,1),\psi_t(x,2),\psi_t(x,3), \psi_t(x,4) )^T \nonumber \\
=\int_0^{2\pi}\{e^{it[\varphi(k)-\frac{x}{t}k]}\langle V_1,\widehat{\Psi_0^{ec}}\rangle V_1\nonumber\\
+e^{it\theta}e^{-ixk}\langle V_2,\widehat{\Psi_0^{ec}}\rangle V_2\nonumber\\
+e^{it\theta}e^{-ixk}\langle V_3,\widehat{\Psi_0^{ec}}\rangle V_3\nonumber\\
+e^{it[2\theta-\varphi(k)-\frac{x}{t}k]}\langle V_4,\widehat{\Psi_0^{ec}} \rangle V_4\} \frac{dk}{2\pi}. \quad \label{eqnprodis3}
\end{eqnarray}

Note that $\frac{d}{dk}(\varphi(k)-\frac{x}{t}k)<\varphi^{\prime}(k)-\varphi^{\prime}(k_0)-\epsilon \leq -\epsilon$, and $\frac{d}{dk}(2\theta-\varphi(k)-\frac{x}{t}k)<-\varphi^{\prime}(k)-\varphi^{\prime}(k_0)-\epsilon<-\epsilon$. Thus, neither $\varphi(k)-\frac{x}{t}k$ nor $2\theta-\varphi(k)-\frac{x}{t}k$ in the first and fourth terms of Eq. (\ref{eqnprodis3}) possess stationary point(s) of non-zero order. By the method of stationary phase in \cite{NV00, Eedelyi56}, $I_j\sim O(t^{-1})$ as $t\rightarrow \infty$, where $j=1,4$. For the second and third terms in Eq. (\ref{eqnprodis3}), by appealing to Lemma 1, we have $\psi_t(x,j)\sim O(x^{-1})$ as $t\rightarrow \infty$ (therefore as $x\rightarrow \infty$), so $I_j\sim O(t^{-1})$ where $j=2,3$. Hence $p_t(x)=O(t^{-2})$. 

(iv)\, \, It suffices to treat the case when $\sqrt{t}<x<(\varphi^{\prime}(k_0)-\epsilon)t$, as the other case can be treated similarly.
Note that $\frac{d}{dk}(\varphi(k)-\frac{x}{t}k)=\varphi^{\prime}(k)-\frac{x}{t}$. since $-M+\epsilon<\frac{x}{t}<M-\epsilon$, we see that $\varphi(k)-\frac{x}{t}k$ and $2\theta-\varphi(k)-\frac{x}{t}k$ both possess a stationary point of order 1. Thus, by the method of stationary phase in \cite{Eedelyi56}, $I_j\sim O(t^{-\frac{1}{2}})$, where $j=1,4$. Meanwhile, the other two terms in Eq. (\ref{eqnprodis3}) satisfy $I_j\sim O(x^{-1})$ where $j=2,3$. Note that $\sqrt{t}<x<(\varphi^{\prime}(k_0)-\epsilon)t$, so either $I_j\sim O(t^{-\frac{1}{2}})$ or $I_j\sim o(t^{-\frac{1}{2}})$ for $j=2,3$. Hence $p_t(x)=O(t^{-1})$.

(v)\,\, As in the proof of (iv), we have $I_j\sim O(t^{-\frac{1}{2}})$ for $j=1,4$. Note that $I_j\sim O(1)$ as $|x|\sim 0$ for $j=2,3$. Therefore

\begin{eqnarray}
p_t(x)&=&\left\|\int_0^{2\pi}\!\!e^{-ixk}\sum_{j=2}^3\langle V_j(k),\widehat{\Psi_0^{ec}}(k)\rangle V_j(k) \frac{dk}{2\pi}\right\|^2\!\!+O(t^{-\frac{1}{2}})\nonumber \\
&=&p(x)+O(t^{-\frac{1}{2}}).
\end{eqnarray}

(vi)\,\, As in the proof of (iv), it can be shown that $I_j\sim O(t^{-\frac{1}{2}})$ for $j=1,4$. Note that $I_j\sim O(|x|^{-1})\sim O(t^{-\frac{1}{2}})$ as $|x|$ approaches $t^{\frac{1}{2}}$ for $j=2,3$, by Lemma 1. Therefore, $p_t(x)=O(t^{-1})$.

\bibliography{apssamp}

\end{document}